\documentclass[journal ]{new-aiaa}
\usepackage[utf8]{inputenc}
\usepackage{textcomp}

\usepackage{graphicx}
\usepackage{amsmath}
\usepackage[version=4]{mhchem}
\usepackage{siunitx}
\usepackage{longtable,tabularx}
\usepackage{pdflscape}
\usepackage{graphicx}
\usepackage{amsmath}
\usepackage{algorithm}
\usepackage{algorithmic}
\usepackage{array}
\usepackage{multirow}
\usepackage{float}
\usepackage{longtable}
\usepackage{hyperref}
\usepackage{cleveref}
\usepackage{graphicx}
\usepackage{caption}
\usepackage{subcaption}
\usepackage{tikz}
\usepackage{mwe}
\usepackage{rotating}

\usetikzlibrary{shapes.geometric, arrows.meta, positioning}
\usepackage[most]{tcolorbox}
\usepackage{gensymb}

\title{An Analytical Methodology for Quantifying Airspace Conflict Rate and Complexity}

\author{Andrew P. Kendall \footnote{Postdoctoral Fellow, Department of Aerospace Engineering and Engineering Mechanics} and John-Paul Clarke\footnote{Ernest Cockrell Jr. Memorial Chair in Engineering, Department of Aerospace Engineering and Engineering Mechanics, AIAA Fellow}}
\affil{The University of Texas at Austin, Austin, TX, 78712-1221}

\begin{document}

\maketitle

\begin{abstract}
Air traffic growth, advanced air mobility, and increasingly autonomous operations are driving the need for scalable and adaptive airspace design methodologies. Central to this challenge is quantifying how traffic flow structure and demand, governed in part by airspace geometry, influence conflict generation and operational complexity. This paper presents an analytical framework for computing conflict rate and conflict probability in structured airspace using stochastic flow models. Traffic streams are modeled as renewal processes with prescribed inter-arrival time distributions, while interactions between flows are captured through geometry-dependent minimum spacing constraints at merges and crossings. Within this formulation, closed-form upper bounds on the expected conflict rate and conflict probability per aircraft are derived as functions of flow configuration and demand. These metrics are interpreted as complementary measures of airspace complexity, reflecting controller workload and per-aircraft operational risk. The methodology is applied to representative hexagonal cell geometries with varying routing structures and flow distributions. Results reveal non-monotonic tradeoffs between routing flexibility, capacity, and conflict generation, with intermediate flow configurations outperforming both highly constrained and highly distributed cases. The proposed framework provides a tractable tool for evaluating airspace design alternatives and complexity-informed traffic management strategies.

\end{abstract}

\section{Introduction \label{sec: introduction}}

Increasing demand for air transportation, as well as the emergence of advanced air mobility (AAM), unmanned aircraft systems (UAS), and higher levels of autonomy, are placing unprecedented pressure on the structure and management of controlled airspace \cite{thipphavong2018urban}. Traditional sector-based designs and procedural separation standards, which were developed for relatively predictable and homogeneous traffic flows, are increasingly challenged by heterogeneous vehicle performance, flexible routing, and dynamic operational constraints \cite{yang2024review}. These trends necessitate more adaptive and scalable approaches to airspace design, in which traffic flows, routing structures, and separation strategies can be systematically configured to balance throughput, safety, and controller workload \cite{levitt2023uam}. Central to this challenge is the need for quantitative measures of airspace complexity that capture the interaction between flow geometry, traffic demand, and separation requirements \cite{abdellaoui2023building}. Such measures are essential for anticipating conflict generation, evaluating design alternatives, and enabling the development of airspace configurations that mitigate conflicts while maintaining operational efficiency.

To address this need, this paper develops an analytical framework for quantifying conflict generation in structured airspace based on stochastic flow models. Traffic streams are modeled as renewal processes with prescribed inter-arrival time distributions, while interactions between flows are characterized through geometry-dependent minimum spacing constraints at merges and crossings. Within this formulation, conflict occurrence is determined by the relative timing between aircraft in intersecting flows, allowing both the expected rate of conflict events and the probability of conflict per aircraft to be expressed in terms of inter-arrival statistics and flow configuration. This approach enables direct evaluation of how routing geometry, flow distribution, and total demand influence conflict dynamics, without reliance on computationally intensive simulation. The resulting framework provides a tractable foundation for exploring design tradeoffs and identifying airspace configurations that balance flexibility, capacity, and conflict risk. To place the proposed framework in context, we briefly review prior work in conflict modeling, airspace complexity assessment, and traffic flow analysis.

Aircraft conflict detection and probability modeling are crucial in ensuring safe and efficient air traffic management. Pairwise conflict models, such as the one proposed by \cite{zhang2022multivariate}, define a collision region between two aircraft and calculate the instantaneous collision probability based on the relative position and uncertainty of their trajectories. These models can be used to predict the likelihood of conflict between two aircraft and provide a basis for conflict detection. Stochastic trajectory uncertainty, which accounts for the randomness and unpredictability of aircraft motion, is a critical aspect of conflict modeling. Brownian motion, as discussed in \cite{ren2024conflict}, is a popular approach to modeling stochastic uncertainty, as it provides a probabilistic description of aircraft motion and enhances the accuracy of flight conflict prediction. To estimate conflict probability, researchers have employed various methods, including Markov chain approximation, reachability computation, and Gaussian process regression \cite{cerezo2022stochastic}. These methods can be used to predict the probability of conflict between two aircraft and provide a basis for conflict detection and resolution.In addition to pairwise conflict models, researchers have also explored multivariate combined collision detection (MCCD) methods, which combine the concept of velocity obstacle method, potential conflict pool, trajectory prediction, and collision probability \cite{zhang2022multivariate}. MCCD methods can be used to detect conflicts between multiple aircraft and provide a more comprehensive understanding of conflict probability.

Airspace complexity and controller workload are crucial factors in air traffic management, and various metrics have been proposed to measure them. Dynamic density (DD) metrics,consider both traffic density and traffic complexity to estimate task demands in real-time. These metrics are widely used in the aviation domain, but they have limitations, such as multi-collinearity among factors and the need for subjective weights \cite{pang2023air}. Other metrics, such as the pairwise dynamic workload (PDW), have been proposed to capture the continuous impact of pairwise operations on airspace structure and controller workload. The PDW is a continuous function that depends on the aircraft separation and separation rate, and it has been shown to be more effective than DD in predicting airspace complexity \cite{souza2026airspace}. Conflict counts and density are also important metrics in air traffic management. Conflict density is a measure of the number of conflicts in a given airspace, and it can be used to estimate controller workload \cite{pang2023air}. However, conflict density metrics have limitations, such as the need for subjective weights and the difficulty in reproducing some of the metrics due to insufficient definition \cite{goppel2024complexity}. Airspace complexity and controller workload are complex phenomena that require a multifaceted approach to measurement. While dynamic density metrics are widely used, they have limitations, and other metrics, such as the pairwise dynamic workload, may be more effective in certain scenarios.

Air traffic flow management is a complex task that involves managing the flow of aircraft through the airspace to ensure safe and efficient flight operations. Several approaches have been employed to model and analyze air traffic flow, including traffic flow/network models, flow-based models, sector capacity, and queueing/network approaches. Traffic flow/network models are used to represent the movement of aircraft through the airspace, taking into account factors such as sector capacity, airspeed, and altitude. These models can be used to analyze the impact of various factors on air traffic flow, such as weather conditions, air traffic control decisions, and system failures \cite{yang2019network}. Flow-based models, on the other hand, focus on the flow of traffic through specific sectors or areas of the airspace \cite{zhang2025short}. These models can be used to predict air traffic flow based on historical data and real-time information, and to identify areas of congestion and potential bottlenecks. Sector capacity is a critical factor in air traffic flow management, as it determines the number of aircraft that can be safely handled within a sector at a given time. Queueing/network approaches are used to model the flow of traffic through sectors, taking into account factors such as sector capacity, airspeed, and altitude \cite{yang2025multi}. Queueing/network models can be used to analyze the impact of various factors on air traffic flow, such as sector capacity, air traffic control decisions, and system failures \cite{tamimi2020cyber}. 

Estimating the rates of separation violations in en route airspace is crucial for air traffic management. Mathematical models using renewal theory and inter-arrival time distributions have been used to analyze and predict separation violation rates. The G/G/c queue model has been used to analyze the time-varying delay time of flights in both en-route and terminal airspace \cite{itoh2020evaluating}. The results show that the control can reduce the mean and maximum delay time for flights by $18.8\%$ and $16.5\%$, respectively, on average within the en-route airspace \cite{higasa2023effectiveness}. Renewal theory has also been applied to estimate separation violation rates. For instance, the authors in \cite{pawelek2019arrival} proposed a method to analyze selected aspects of past arrival traffic by modeling distributions of time separations of arriving aircraft in a chosen navigation point of Terminal Maneuvering Area using continuous probability distributions. The method allows for quantitative analysis of separations for days with various arrival traffic intensity by comparing distributions parameters.

Taken together, these prior efforts highlight the need for an analytical framework that directly links stochastic flow generation, airspace geometry, and conflict dynamics in a form suitable for design and trade studies. The approach developed in this paper addresses this gap by modeling interacting traffic flows as renewal processes and deriving expressions for both conflict rate and conflict probability based on geometry-dependent spacing constraints. This formulation enables systematic evaluation of how routing structure, flow distribution, and total demand influence conflict behavior without reliance on simulation. The primary contributions of this work are:
\begin{enumerate}
\item An analytical framework for computing conflict rate and conflict probability in interacting air traffic flows using renewal theory.
\item A geometry-dependent formulation for crossing and merging conflicts.
\item An evaluation of conflict-based airspace complexity metrics across representative hexagonal routing structures.
\item Design insights regarding the tradeoff between routing flexibility, capacity, and conflict generation.
\end{enumerate}
The remainder of the paper is organized as follows: in Section II we present the analytical derivation of an upper bound conflict rate and probability; in Section III we introduce the modeling framework for geometry dependent minimum separation requirements and inter-arrival time distributions; in Section IV we demonstrate the methodology through representative sector cell geometries and flow configurations; and in Section V we conclude with key findings and implications for airspace design.

\section{Conflict Rate Estimation \label{sec: methodology}}

Our proposed methodology for quantifying airspace complexity requires connecting the arrival rate of aircraft flows and their routing geometry to an estimate of conflict rate for a sector of airspace. In this section we introduce the problem structure and derive upper bounds on conflict rates and probabilities using elementary renewal theory.

\subsection{Conflict Probability}

We define a conflict as a realized loss of separation, i.e., an event in which the distance between two aircraft falls below the minimum required threshold $r_o$. We aim to compute two related metrics: (i) an upper bound on the conflict probability $P_C$, defined as the expected probability that an aircraft passing through a cell experiences at least one conflict, and (ii) an upper bound on the conflict rate $Q_C$, defined as the expected number of conflict events within a cell per unit time. It is assumed that we know the geometry of the different paths involved and that the inter-arrival time $S_i$ for each individual flow $i$ has Probability Density Function (PDF) $f_i(s)$. The geometry of the path and the relative velocity of each pair of flows affect the minimum inter-arrival time $s_{i,j}$ between two flows required to maintain spatial separation. 

Each aircraft starts from an origin flow $o$ with total flow rate $\lambda_o$ and ends merged into a destination flow $d$ with total flow rate $\lambda_d$. Each individual flow $i$ within the cell connects an origin flow $o_i$ to a destination flow $d_i$ with an individual flow rate $\lambda_i$. We assume the system is in a steady state such that $\lambda_o = \sum_{i:o_i=o}\lambda_i$, $\lambda_d = \sum_{i:d_i=d}\lambda_i$.

The aim is to estimate the probability that a random arrival from flow $i$ will experience a conflict with any other flow within the cell. This event, $\mathcal{V}_i$, is a union of the conflict event $\mathcal{V}_{i,o}$ with each origin flow $o$ in the set of origin flows $\mathcal O$.
\begin{equation}
\mathcal{V}_{i} = \bigcup_{o\in\mathcal{O}}\mathcal{V}_{i,o}
\end{equation}

The conflict event between flow $i$ and each combined origin flow is itself the union of the conflict events $\mathcal{V}_{i,k}$ between flow $i$ and each flow $k$ with origin $o$.
\begin{equation}
\mathcal{V}_{i,o} = \bigcup_{k:o_k=o}\mathcal{V}_{i,k}
\label{eq: origin_event}
\end{equation}

It is assumed that each origin flow $o$ entering the cell is already spaced with inter-arrival PDF $f_o(s)$, and is independent of other flows with different origins. The probability of the union of origin flow conflict events may be broken down in terms of the individual independent pairwise event probabilities by De Morgan's laws, this is a standard treatment for a superposition of arrival process \cite{lawrance1973dependency}. 

\begin{equation}
\mathbb{P}(\mathcal{V}_{i}) = \mathbb{P}\bigg(\bigcup_{o\in\mathcal{O}}\mathcal{V}_{i,o}\bigg) = 1-\sum_{o\in\mathcal{O}}\bigg(1-\mathbb{P}(\mathcal{V}_{i,o})\bigg)
\end{equation}

This decomposition results in 2 cases that are treated slightly differently. Case I involves conflict events between flow $i$ and the other flows from its own origin $o_i$ (including itself), while Case II involves the remaining conflicts with flows of different origin.

We first consider pairwise flow conflicts in Case I. Although flows with the same origin start out spaced properly, diverging, curved, and variable speed paths may result in conflicts of previously spaced flows. The probability of a spacing violation is computed using the inter-arrival Cumulative Distribution Function (CDF) $F_j(s)$ and the path geometry dependent minimum separation time $s_{i,j}$ between flow $i$ and $j$. 

\begin{equation}
\begin{aligned}
\mathbb{P}(\mathcal{V}_{i,j}) = \mathbb{P}(S_{j}\leq s_{i,j})  = 1-F_{j}(s_{i,j}), \quad o_j=o_i
\end{aligned}
\label{eq: arrival_phi}
\end{equation}

The probability of the union of such events for each flow from origin $o_i$ demands a more modified treatment to be derived in a later section. This is due to the correlation between arrival times for each flow constituting an origin flow. The pairwise conflicts of Case II cannot be treated the same as in Case I due to each stemming from a different origin flow.

\subsection{Renewal Theory}

Because flows from different origins are uncorrelated, conflict probabilities between pairs of such flows are naturally treated using renewal theory. Each flow $i$ is modeled as a renewal process $N_i(t)$, which counts the number of arrivals up to time $t$. Let $J_k$ denote the time of the k-th arrival, and let the inter-arrival times

\begin{equation}
S_k = J_k-J_{k-1},\quad k\geq1,
\end{equation}
be i.i.d. with PDF $f_i(s)$, CDF $F_i(s)$, and finite mean
\begin{equation}
\mathbb{E}[ S_k] = \lambda_i^{-1}.
\end{equation}
$N_i(t)$ counts the number of arrivals up to time $t$ as 
\begin{equation}
N_i(t) = \sup\{k:J_k\leq t\}, \quad N_i(0)=0.
\end{equation}
The residual time at time t is defined as 
\begin{equation} R_i(t) = J_{N_i(t)+1}-t,
\end{equation}
representing the time from 
$t$ until the next arrival. In the aircraft spacing context, this residual time corresponds to the separation between the closest aircraft in two uncorrelated flows.

Rather than conditioning on a specific observation time 
$t$, we are interested in the stationary residual lifetime, defined as the limiting distribution of $R_i(t)$ as $t\rightarrow \infty$. Under the assumption that $\mathbb{E}[S_k]\leq \infty$, the elementary renewal theorem implies that the renewal process admits a stationary version, and that the distribution of $R_i(t)$ converges to a limiting distribution independent of $t$ \cite{feller1991introduction}.

For $s\geq 0$, the stationary residual lifetime distribution satisfies

\begin{equation}
\mathbb{P}(R_i>s) = \lim_{t\rightarrow \infty} \mathbb{P}(R_i(t)>s).
\end{equation}

By renewal theory, the probability that a randomly chosen time falls within an inter-arrival interval of length $u$ is proportional to $uf_i(u)$. Conditioning on the length of the interval containing the observation time, we obtain

\begin{equation}
\mathbb{P}(R_i>s) = \frac{1}{\mathbb{E}[S_k]}\int_s^\infty \mathbb{P}(S_k>u) du = \lambda_i \int_s^\infty [1-F_i(u)] du.
\end{equation}

Differentiating with respect to 
$s$ yields the stationary residual lifetime PDF

\begin{equation}
\begin{aligned}
\phi_i(s)&= \lambda_i [1-F_i(s)],
\end{aligned}
\label{eq: arrival_phi}
\end{equation}
 and the corresponding complementary CDF
\begin{equation}
\begin{aligned}
\Phi_i(s)=\mathbb{P}(R_i> s)= \lambda_i\int_s^\infty \big[1-F_i(x)\big]dx.
\end{aligned}
\label{eq: arrival_Phi}
\end{equation}

We can now compute the pairwise conflict probability of Case II, for a pair of flows with different origins and path geometry dependent minimum separation time $s_{i,j}$.

\begin{equation}
\begin{aligned}
\mathbb{P}(\mathcal{V}_{i,j}) = \mathbb{P}(R_{j}\leq s_{i,j})  = 1-\Phi_{j}(s_{i,j}), \quad o_j\neq o_i
\end{aligned}
\label{eq: arrival_phi}
\end{equation}

\subsection{Origin Flow Conflict Probability}

Although we have the pairwise conflict probabilities for both Case I and Case II, an origin flow event $\mathcal{V}_{i,o}$ as given in Eq.\ref{eq: origin_event} requires the union of the pairwise events corresponding to a given origin flow $o$. The arrival times of aircraft in each individual flow are not independent, as a minimum spacing is assumed between them. The arrival process is best modeled by sampling arrival times from the combined origin flow, then assigning each arrival to an individual flow with a probability corresponding to its fraction of the total origin flow. Due to the correlated arrival times, we must also consider a time offset $\tau_{i,k}$ between the arrival of time of an aircraft in flow $i$, and the conflict time window of duration $s_{i,k}$ with respect to flow $k$.

We compute an upper bound on the probability of a conflict between flow $i$ and the origin flow $o_j$ using the inclusion-exclusion principle. After considering each individual flow independently, we estimate the probability that at least 1 arrival from flow $k$ occurs between time $\tau_{i,k}$ and $\tau_{i,k}+s_{i,k}$ relative to a random arrival from flow $i$. Given that the combined origin flow inter-arrival spacing distribution is known, we can compute the probability that $n$ arrivals occur during a critical time window, multiply by the probability that at least 1 arrival of $n$ was part of individual flow $k$, and sum over all numbers of arrivals for each individual flow. By treating each individual flow independently, we may neglect the time offset $\tau_{i,k}$.


\begin{equation}
\begin{aligned}
\mathbb{P}(\mathcal{V}_{i,o_j})&=\mathbb{P}\left(\bigcup_{k:o_k=o_j}\mathcal{V}_{i,k}\right)\\&\leq \sum_{k:o_k=o_j}\mathbb{P}(\mathcal{V}_{i,k})  
\\&\leq \sum_{k:o_k=o_j} \mathbb{P}(N_{k}(\tau_{i,k}+s_{i,k})-N_{k}(\tau_{i,k})\geq 1)  
\\&\leq \sum_{k:o_k=o_j}\sum_{n=1}^\infty \left(1-\left(1-\frac{\lambda_k}{\lambda_{o_j}}\right)^n\right) \mathbb{P}(N_{o_j}(\tau_{i,k}+s_{i,k})-N_{o_j}(\tau_{i,k})=n)
\\&\leq \sum_{k:o_k=o_j}\sum_{n=1}^\infty \left(1-\left(1-\frac{\lambda_k}{\lambda_{o_j}}\right)^n\right) \mathbb{P}(J_{o_j,n}\leq s_{i,k},J_{o_j,n+1}> s_{i,k})
\end{aligned}
\label{eq: origin_sep_violation}
\end{equation}

We consider Case I first, in which the the origin of flow $i$ is origin flow $o_j$, and the first arrival time, $J_0$, is sampled from the inter-arrival time distribution.

\begin{equation}
\begin{aligned}
J_n=\sum_{i=0}^nS_i, \quad o_i= o_j
\end{aligned}
\label{eq: case1Jn}
\end{equation}

Using Eq.\eqref{eq: origin_sep_violation} and results from Appendix I, we compute the first order approximation of the Case I conflict probability between flow $i$ and origin flow $o_i$. For scenarios of interest in this study, $s_{i,k}<2s_0$ for Case I flow pairs, so higher order terms in the approximation may be neglected.

\begin{equation}
\begin{aligned}
\mathbb{P}(\mathcal{V}_{i,o_j}|o_i = o_j)
&\leq \sum_{k:o_k=o_j}\bigg[\frac{\lambda_k}{\lambda_{o_j}}F_{o_j}(s_{i,k})\\&\quad+\sum_{n=1}^{\lfloor\frac{s_{i,k}}{s_0} \rfloor} \left(1-\left(1-\frac{\lambda_k}{\lambda_{o_j}}\right)^{n+1}\right) F_{o_j}^{n+1}(s_{i,k}-ns_0)\bigg]
\\
&\lessapprox \sum_{k:o_k=o_j}\frac{\lambda_k}{\lambda_{o_j}}F_{o_j}(s_{i,k}) \\
\end{aligned}
\label{eq: arrival_CDF_different_same}
\end{equation}

For Case II, in which the origin of flow $i$ is $\textit{not}$ origin flow $o_j$, the first arrival time, $J_0$, is sampled from the residual lifetime time distribution.

\begin{equation}
\begin{aligned}
J_0 = R,\,\, J_{n\geq1}=R+\sum_{i=1}^nS_i, \quad o_i\neq o_j
\end{aligned}
\label{eq: case2Jn}
\end{equation}

\begin{figure}
    \centering
    \includegraphics[width=1.0\linewidth]{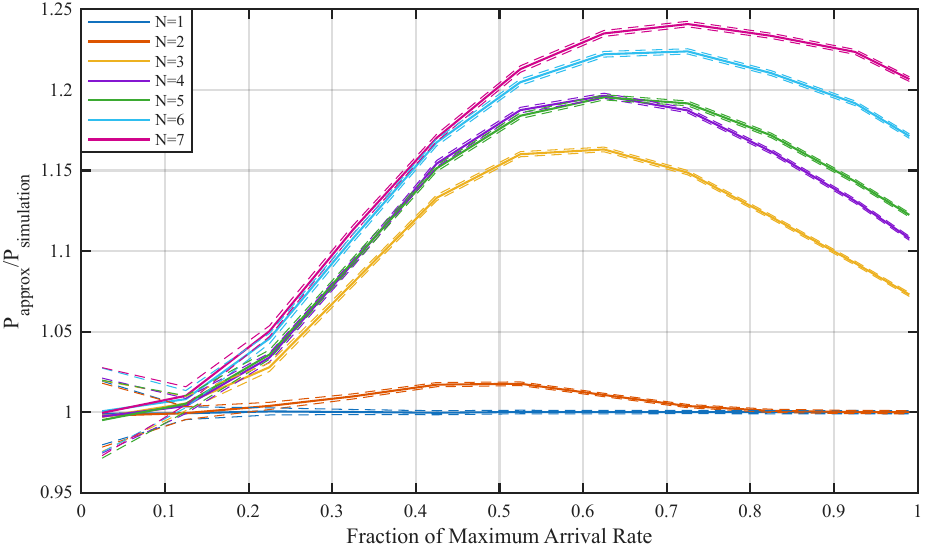}
    \caption{Case II conflict probability approximation ratio against Monte-Carlo discrete event simulation estimate for a range of interacting flows with $2\sigma$ confidence intervals.}
    \label{fig:CASE2_err}
\end{figure}

Using a similar method to Case I, we compute the first order approximation of the Case II conflict probability between flow $i$ and origin flow $o_j \neq o_i$.

\begin{equation}
\begin{aligned}
\mathbb{P}&(\mathcal{V}_{i,o_j}|o_i\neq o_j)
\leq \sum_{k:o_k=o_j}\bigg[\frac{\lambda_k}{\lambda_{o_j}}(1-\Phi_{o_j}(s_{i,k}))\\&+\sum_{n=1}^{\lfloor\frac{s_{i,k}}{s_0} \rfloor} \left(1-\left(1-\frac{\lambda_k}{\lambda_{o_j}}\right)^{n+1}\right) F_{o_j}^n(s_{i,k}-(n-1)s_0)(1-\Phi_{o_j}(s_{i,k}-n s_0))\bigg]
\\
&\lessapprox \sum_{k:o_k=o_j}\frac{\lambda_k}{\lambda_{o_j}}(1-\Phi_{o_j}(s_{i,k})) \\
\end{aligned}
\label{eq: sep_prob_case2}
\end{equation}
 Neglecting of higher order terms is found to be a conservative approximation for scenarios of interest in this study. We perform a Monte-Carlo estimation of Case II conflict probabilities using discrete event simulation and representative scenario parameters. Figure \ref{fig:CASE2_err} shows the ratio of the approximation given in Eq.\eqref{eq: sep_prob_case2} to the simulated probability for a range of arrival rates and the number of individual flows constituting the origin flow. The approximation is accurate for low arrival rates and few flows, but overestimates by roughly $24\%$ in the worst case, due to treating each flow independently and neglecting even order terms in the inclusion-exclusion expansion.

\subsection{Total Conflict Rate Estimate}

The previous section allows us to compute upper bounds on the total conflict probability for flow $i$.

\begin{equation}
\begin{aligned}
P(\mathcal{V}_i) \lessapprox 1 - \left(1-\sum_{k:o_k=o_j}\frac{\lambda_k}{\lambda_{o_i}}F_{o_i}(s_{i,k})\right) \left(\sum_{o_j\neq o_i}\sum_{k:o_k=o_j}\frac{\lambda_k}{\lambda_{o_j}}(1-\Phi_{o_j}(s_{i,k}))\right)
\end{aligned} 
\label{eq: coll_rate_i}
\end{equation}

We define the conflict rate of individual flow $i$ as the conflict probability upper bound multiplied by the arrival rate $\lambda_i$. 

\begin{equation}
\begin{aligned}
Q_i =\lambda_i - \lambda_i\left(1-\sum_{k:o_k=o_j}\frac{\lambda_k}{\lambda_{o_i}}F_{o_i}(s_{i,k})\right) \left(\sum_{o_j\neq o_i}\sum_{k:o_k=o_j}\frac{\lambda_k}{\lambda_{o_j}}(1-\Phi_{o_j}(s_{i,k}))\right)
\end{aligned}
\label{eq: coll_rate_i}
\end{equation}

The conflict rate for $\textit{all}$ flows within a cell is defined as the total sum of all individual flow conflict rates divided by 2 to account for the over-counting of pairs.
\begin{equation}
\begin{aligned}
Q_C= \frac{1}{2}\sum_i Q_i
\end{aligned}
\label{eq: coll_rate}
\end{equation}

$Q_C$ serves as the metric for airspace complexity, as it is associated with the amount of work per unit of time required to resolve conflicts within a cell.
The effective conflict probability per aircraft passing through the cell, $P_C$, is computed by normalizing $Q_C$ by the total flow rate though the cell $\lambda_C=\sum_i\lambda_i$.

\begin{equation}
\begin{aligned}
P_C= \frac{Q_C}{\lambda_C}
\end{aligned}
\label{eq: coll_rate}
\end{equation}

$P_C$ serves as an efficiency metric for airspace complexity, as it balances the work per unit time with airspace throughput.
\section{Separation and Spacing Modeling \label{sec: modelmethodology}}

The conflict rate and probability model derived in the previous section requires several parameters in order to be applied to an airspace configuration. The first set of parameters, $s_{i,j}$, describe the minimum required time separation between each pair of flows, and are dependent upon the routing geometry and kinematics of aircraft flows within a sector of airspace. The second set of parameters describe the inter-arrival time distribution of aircraft in each origin flow, as well as the related residual lifetime distribution.

\subsection{Minimum Separation Time Modeling}

\begin{figure*}[t!]
    \centering
    \begin{subfigure}[t]{0.33\textwidth}
        \centering
        \includegraphics[height=0.85in]{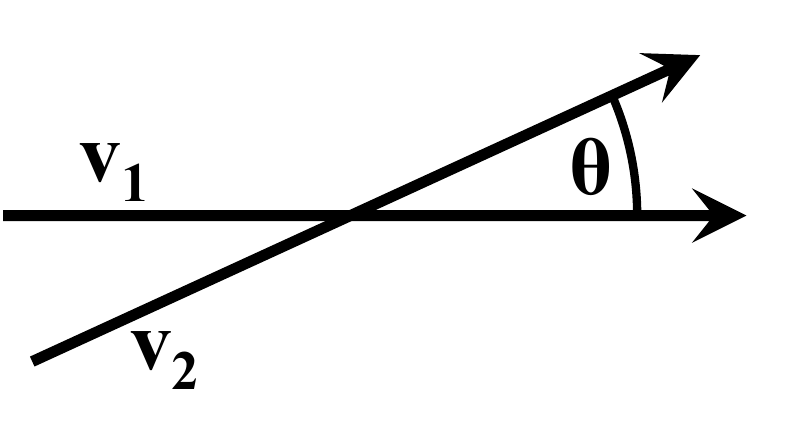}
        \caption{Crossing Geometry}
        \label{fig:cross_geometry}
    \end{subfigure}%
    ~ 
    \begin{subfigure}[t]{0.33\textwidth}
        \centering
        \includegraphics[height=1.2in]{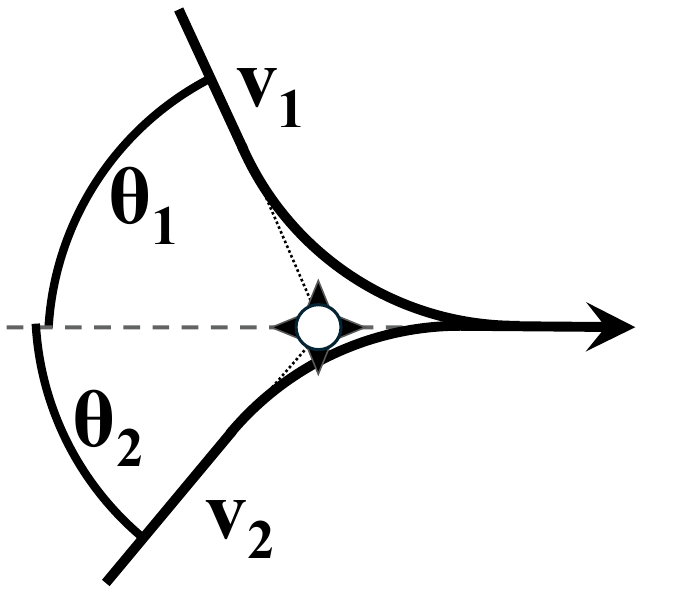}
        \caption{Flyby Merge Geometry}
        \label{fig:flyby_geometry}
    \end{subfigure}
     ~ 
    \begin{subfigure}[t]{0.33\textwidth}
        \centering
        \includegraphics[height=1.2in]{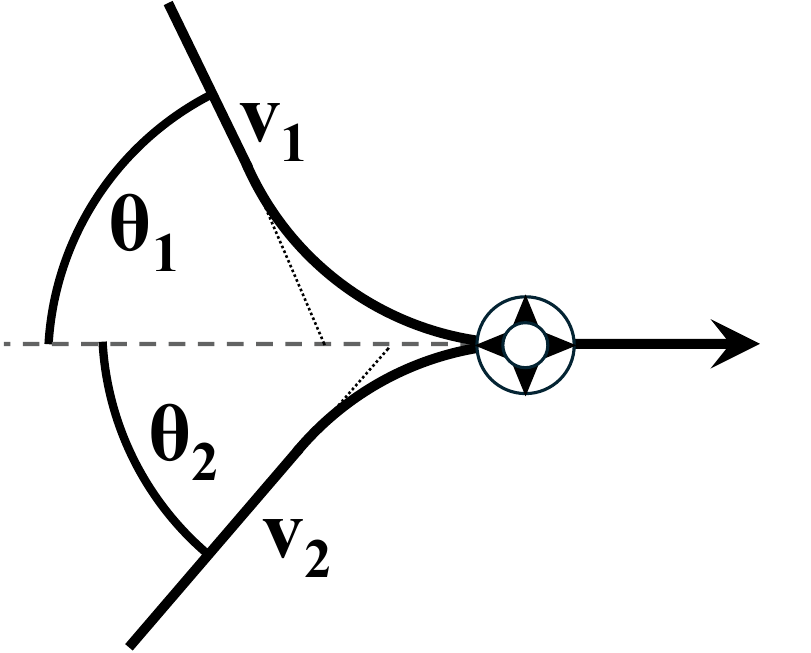}
        \caption{Flyover Merge Geometry}
        \label{fig:flyover_geometry}
    \end{subfigure}
    \caption{Aircraft Path Interaction Geometries}
    \label{fig:merge_geometry}
\end{figure*}

Pairs of aircraft on the same path at constant speed will have the required minimum inter-arrival time spacing $s_{min}=\frac{r_{0}}{v}$, however, differing velocities and path geometries between the two aircraft will change the required minimum spacing. Path pair geometries that can lead to conflicts include crossings, merges, and divergences. Transient merges occur when two paths temporarily merge within a cell before diverging again. Proximity conflicts may occur when two paths do not intersect, but come within the minimum required spacing. 
We can analytically compute the minimum required time separation $s$ of two straight paths crossing at angle $\theta$ with velocities $v_1$ and $v_2$. The geometry of the crossing is depicted in figure \ref{fig:cross_geometry}.

\begin{equation}
\begin{aligned}
x_1 &= v_1 t \\
y_1 &= 0 \\
x_2 &= v_2 \cos(\theta) (t-s) \\
y_2 &= v_2 \sin(\theta) (t-s) \\ \\
\end{aligned}
\label{eq: xyspacing}
\end{equation}

\begin{figure}
    \centering
    \includegraphics[width=0.95\linewidth]{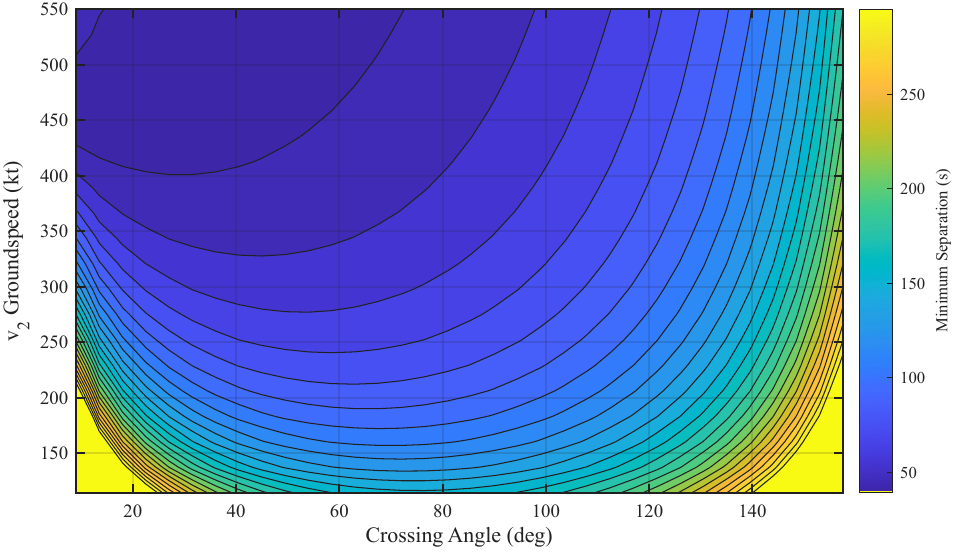}
    \caption{Straight crossing effective minimum separation, $v_1 = 458.4$ kt, $r_0=5$ NM.}
    \label{fig:crossingspace}
\end{figure}

We find the spacing $s$ such that the minimum distance $d_{12}$ between the aircraft is equal to $r_0$. Equation\eqref{eq: crossing_spacing} is plotted for parameters of interest in figure \ref{fig:crossingspace}.

\begin{equation}
\begin{aligned}
d_{12}(t,s) = \sqrt{\big(x_1(t)-x_2(t,s)\big)^2+\big(y_1(t)-y_2(t,s)\big)^2}
\end{aligned}
\label{eq: d12spacing}
\end{equation}

\begin{equation}
\begin{aligned}
s_{12} = \{s:\min_t d_{12}(t,s) = r_0\}
\end{aligned}
\label{eq: smin}
\end{equation}

\begin{equation}
s_{12,\,cross} = r_o \frac{\sqrt{v_1^2+v_2^2-2v_1v_2\cos\theta}}{v_1v_2}
\label{eq: crossing_spacing}
\end{equation}

\begin{figure}
    \centering
    \includegraphics[width=0.95\linewidth]{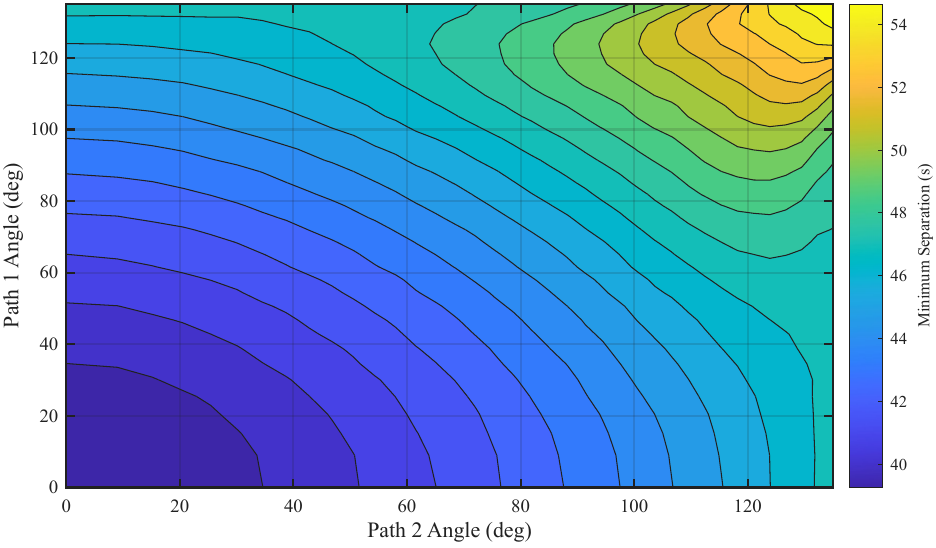}
    \caption{Flyby merge effective minimum separation,  $v_1=v_2 = 458.4$ kt, $r_0=5$ NM, $\dot\psi=3 \degree s^{-1}$.}
    \label{fig:flybyspace}
\end{figure}

\begin{figure}
    \centering
    \includegraphics[width=0.95\linewidth]{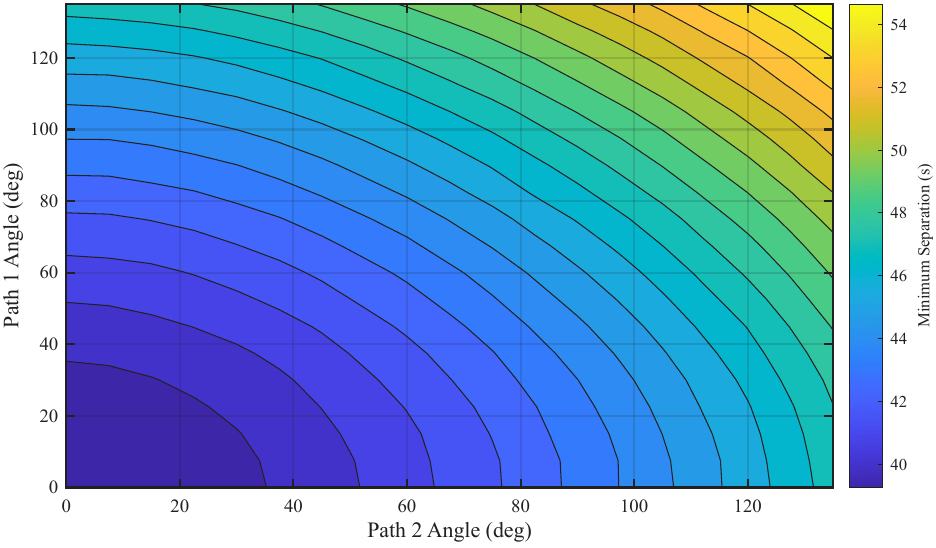}
    \caption{Flyover merge effective minimum separation, $v_1=v_2 = 458.4$ kt, $r_0=5$ NM, $\dot\psi=3 \degree s^{-1}$.}
    \label{fig:flyoverspace}
\end{figure}

Other path pair geometries such as merges do not admit a general analytic solution and are computed numerically. A line search is used to find the mean of the lead and trailing spacing of one trajectory with respect to another that correspond to minimum distances of $r_0$. It is assumed that paths consist of straight line segments connected by arcs with radii corresponding to a turn rate, $\dot\psi$, of $3 \degree s^{-1}$ . The angle dependent spacing for two particular merge geometries, flyby and flyover, are computed. Flyby merge geometries, depicted in figure \ref{fig:flyby_geometry}, consist of straight incoming paths aimed at the flyby merge point which round off the corner to join the merged outward path. Minimum flyby merge spacing for parameters of interest is shown in figure \ref{fig:flybyspace}. Flyover merge geometries, depicted in figure \ref{fig:flyover_geometry}, require paths to aim behind a flyover merge point and perform the turn earlier in order to merge with the outgoing path exactly over the merge point. Minimum flyover merge spacing for parameters of interest is shown in figure \ref{fig:flyoverspace}. The differences between the two cases are subtle, but produce dissimilar minimum spacing behavior at merge angles over $90\degree$.

\subsubsection{Inter-arrival Time Modeling}

The inter-arrival times are modeled with a shifted Log-Logistic distribution. This distribution provides a flexible, heavy-tailed model for inter-arrival times that captures both minimum spacing effects and variability in upstream flow generation, while remaining analytically tractable for convolution and residual-life calculations. The parameter $s_{min}$ sets the minimum inter-arrival time of a flow and $\lambda$ is the total flow rate. The $\alpha$ and $\beta$ parameters of the un-shifted Log-Logistic distribution are associated with the scale and shape, respectively, and are computed as a function of $\lambda$ and $s_{min}$.

\begin{equation}
\begin{aligned}
f(s;s_{min},\alpha,\beta) = \mathbf{1}[s\geq s_{min}]\frac{\beta}{\alpha} \frac{(\frac{s-s_{min}}{\alpha})^{\beta-1}}{\big(1+(\frac{s-s_{min}}{\alpha})^{\beta}\big)^2}
\end{aligned}
\label{eq: fLL}
\end{equation}

\begin{equation}
\begin{aligned}
F(s;s_{min},\alpha,\beta) = \mathbf{1}[s\geq s_{min}] \frac{1}{1+(\frac{s-s_{min}}{\alpha})^{-\beta}}
\end{aligned}
\label{eq: FLL}
\end{equation}


The free shape parameter $\beta$ is found to be $3.052$ by applying maximum likelihood estimation to observed spacing distributions in terminal areas \cite{ren2007separation}. This also ensures that the first three moments of the distribution are finite.
The residual lifetime PDF and complementary CDF may be computed for the Log-Logistic distribution,

\begin{equation}
\begin{aligned}
\phi(s;s_{min},\alpha,\beta,\lambda) = \lambda\left(1-\mathbf{1}[s\geq s_{min}] \frac{1}{1+(\frac{s-s_{min}}{\alpha})^{-\beta}}\right)
\end{aligned}
\label{eq: phiLL}
\end{equation}

\begin{equation}
\begin{aligned}
\Phi(&s;s_{min},\alpha,\beta,\lambda) = \lambda\bigg(\lambda^{-1}-s+\mathbf{1}[s< s_{min}] (s-t)\\ &+\mathbf{1}[s\geq s_{min}] \left[(s-t)\,\, _2F_1\left(1,\beta^{-1};1+\beta^{-1};-\left(\frac{t-s}{\alpha} \right)^\beta\right)\right]\bigg),\\
\end{aligned}
\label{eq: phiLL}
\end{equation}

where $ _2F_1$ is the Hypergeometric function \cite{DLMF}.

\section{Results \label{sec: numerical_study}}

The foundation of this study is a sector modelled as a unit grid element or $\textit{cell}$ of uniform height that is hexagonal in plan with side lengths three times the separation minima. Following the conventional East-West traffic flow routing by flight level, we constrain our study to scenarios with flows traveling at constant altitude with headings greater than $0\degree$ and less than $180\degree$. Four cell geometries are considered, cell geometries 1 and 2, with the 3 origin flows and 3 destination flows, and cell geometries 3 and 4, with 7 origin flows and 7 destination flows. Geometry 1 is a modification of geometry 3 reduced to only the 3 origin and destination flows that are perpendicular to the sides of the cell. Similarly, geometry 2 is also produced by the same modification applied to geometry 4. Cell geometries 1 and 3 have more direct routing from origins to destinations, while geometries 2 and 4 have a fly-by way-point in the center of the cell. We consider constant ground speeds of $458.4$ knots, turn rates $\dot\psi=3 \degree s^{-1}$, en-route separation minimum $r_0=5$ NM, and cell side lengths of $15$ NM. Visualizations of each cell routing geometry as well as the minimum time separation $s_{i,j}$ between each pair of flows are presented in Appendix II.  

\begin{figure*}[t!]
    \centering
    \begin{subfigure}[t]{0.45\textwidth}
        \centering
        \includegraphics[height=1.9in]{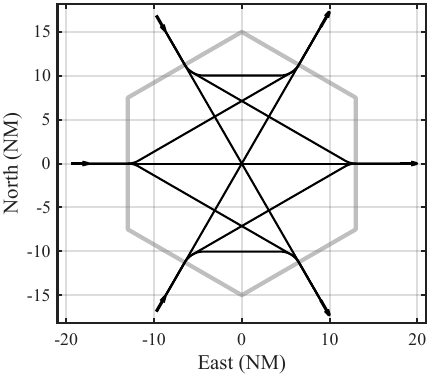}
        \caption{Cell Routing Geometry 1}
        \label{fig:hex1small}
    \end{subfigure}%
        ~ 
    \begin{subfigure}[t]{0.45\textwidth}
        \centering
        \includegraphics[height=1.9in]{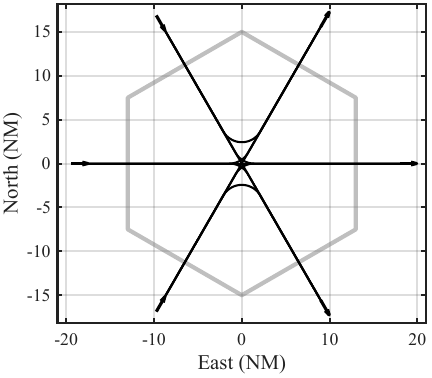}
        \caption{Cell Routing Geometry 2}
        \label{fig:hex2small}
    \end{subfigure}
    ~ 
    ~ 
    \begin{subfigure}[t]{0.45\textwidth}
        \centering
        \includegraphics[height=1.9in]{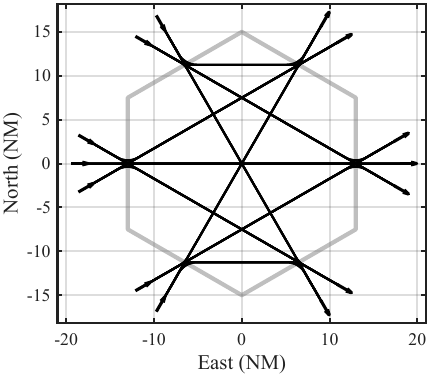}
        \caption{Cell Routing Geometry 3}
        \label{fig:hex3small}
    \end{subfigure}
     ~ 
    \begin{subfigure}[t]{0.45\textwidth}
        \centering
        \includegraphics[height=1.9in]{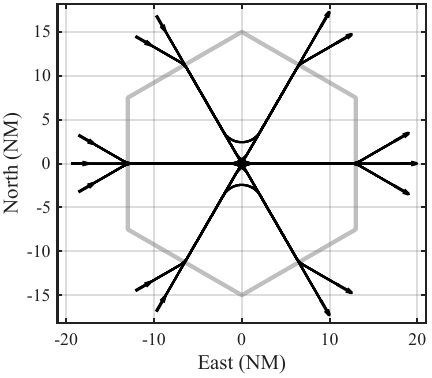}
        \caption{Cell Routing Geometry 4}
        \label{fig:hex4small}
    \end{subfigure}
    \caption{Cell Routing Geometries}
    \label{fig:smallhex}
\end{figure*}

\subsection{Complexity Comparison}

\begin{figure}
    \centering
    \includegraphics[width=0.95\linewidth]{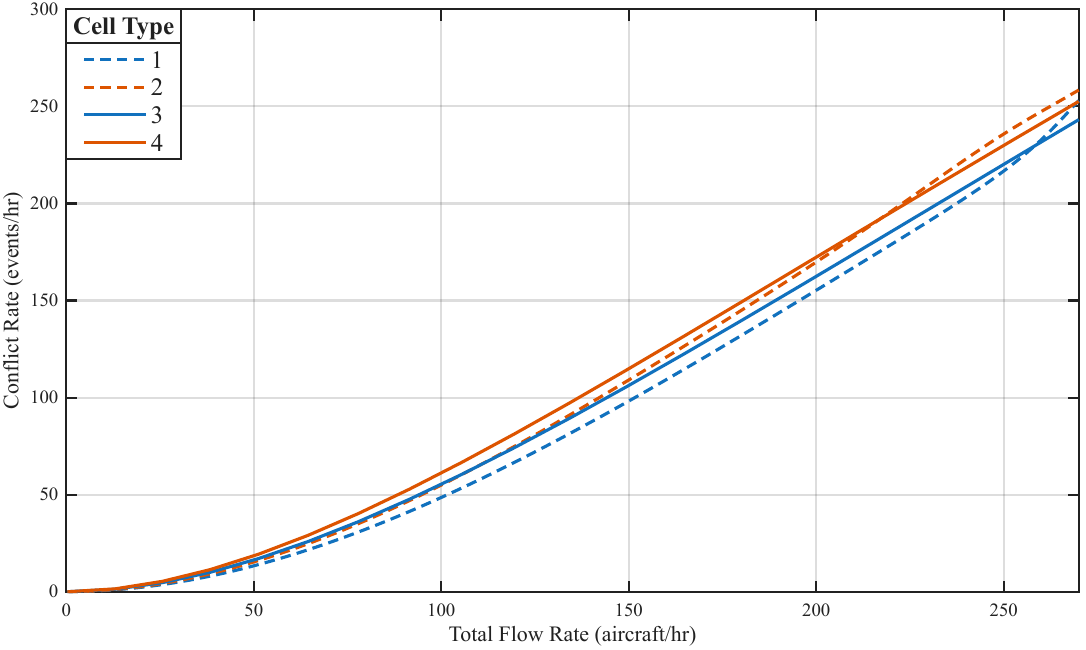}
    \caption{Conflict Rate Metric for each Cell Geometry.}
    \label{fig:rate_compare}
\end{figure}

\begin{figure}
    \centering
    \includegraphics[width=0.95\linewidth]{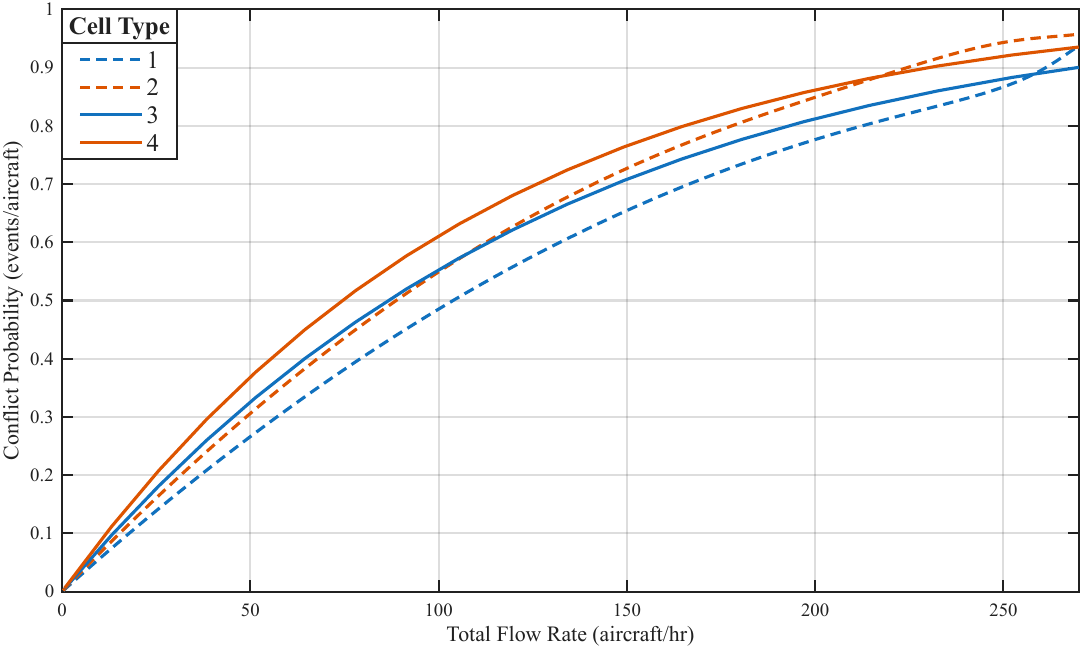}
    \caption{Conflict Probability Metric for each Cell Geometry.}
    \label{fig:prob_compare}
\end{figure}

The conflict rate for each cell type as a function of the total flow rate, equally distributed across all origin flows, is shown in figure \ref{fig:rate_compare}. The corresponding conflict probability, conflict rate normalized by total flow rate, is shown in figure \ref{fig:prob_compare}. It is apparent that for low flow rates, the 3 by 3 Cell Types 1 and 2 have lower conflict probabilities compared to the 7 by 7 Cell Types 3 and 4. This is attributed to the reduction in number of crossings and flow interactions in the 3 by 3 Cells. However, at high flow rates, the 3 by 3 Cells have a larger conflict probability compared to the 7 by 7 Cells. This is due to the reduced capacity of the 3 by 3 Cells, as each  individual flow is closer to its maximum capacity and the origin flows experience self interaction conflicts. Cell Types 1 and 3, with more direct routing, have overall lower conflict probabilities than the corresponding centrally routed Cell Types 2 and 4. This is attributed to the smaller crossing and merge angles in the direct routing Cells, as well as the avoidance of unnecessary flow crowding in the center.

\subsection{Flow Distribution Parameter}

\begin{figure}
    \centering
    \includegraphics[width=0.95\linewidth]{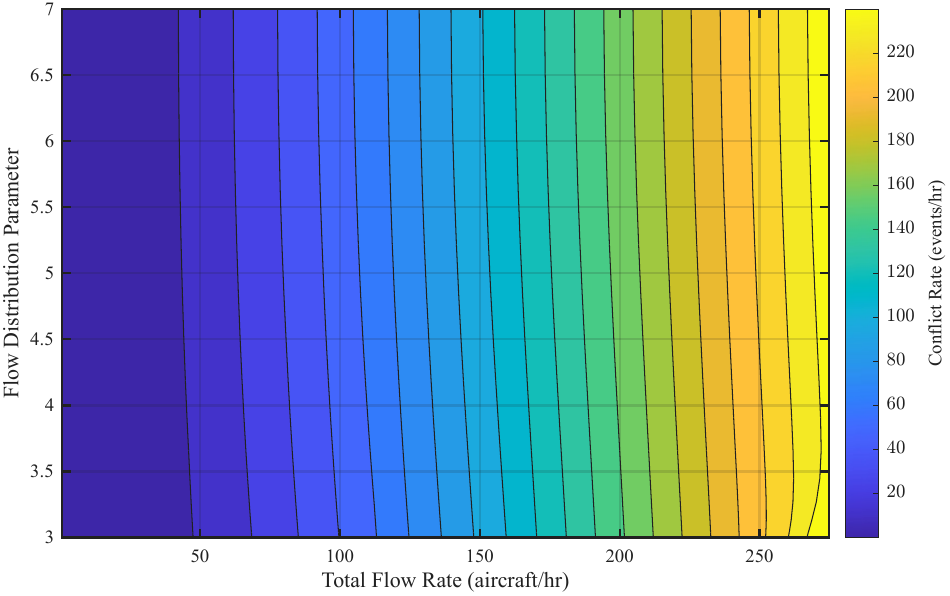}
    \caption{Cell Geometry 3 Conflict Rate as a function of flow rate and flow distribution parameter.}
    \label{fig:Rate_hex13}
\end{figure}

\begin{figure}
    \centering
    \includegraphics[width=0.95\linewidth]{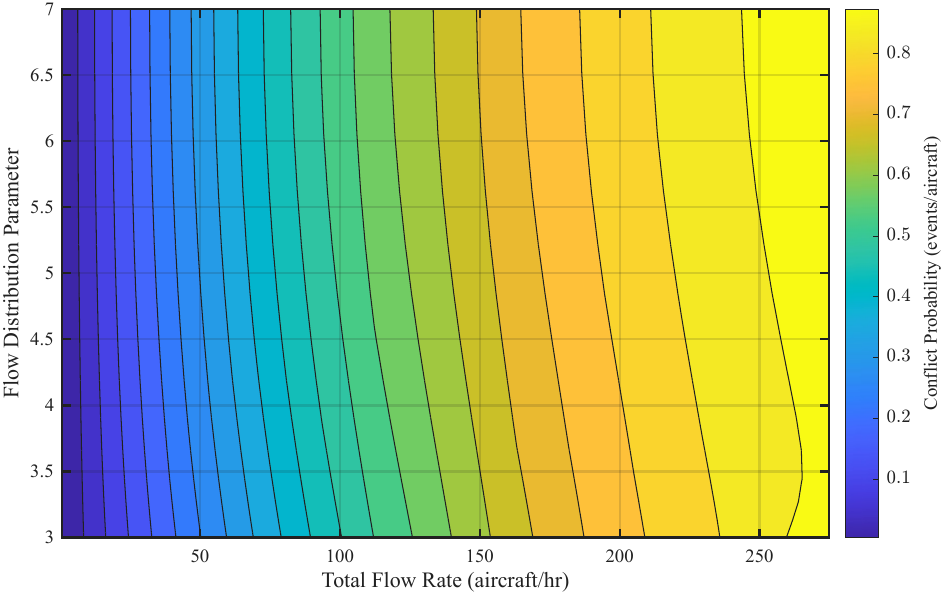}
    \caption{Cell Geometry 3 Conflict Probability as a function of flow rate and flow distribution parameter.}
    \label{fig:PR_hex13}
\end{figure}

\begin{figure}
    \centering
    \includegraphics[width=0.95\linewidth]{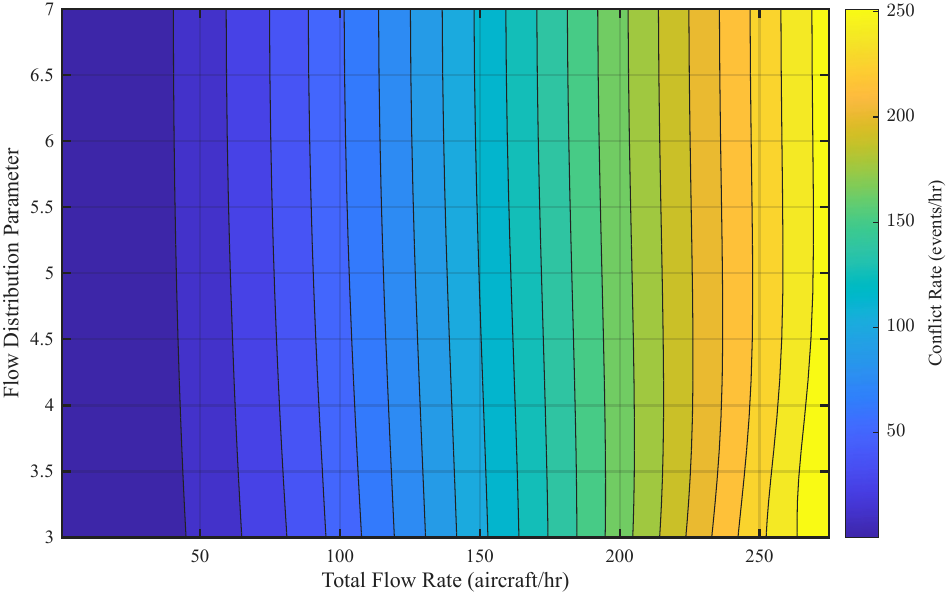}
    \caption{Cell Geometry 4 Conflict Rate as a function of flow rate and flow distribution parameter.}
    \label{fig:Rate_hex24}
\end{figure}

\begin{figure}
    \centering
    \includegraphics[width=0.95\linewidth]{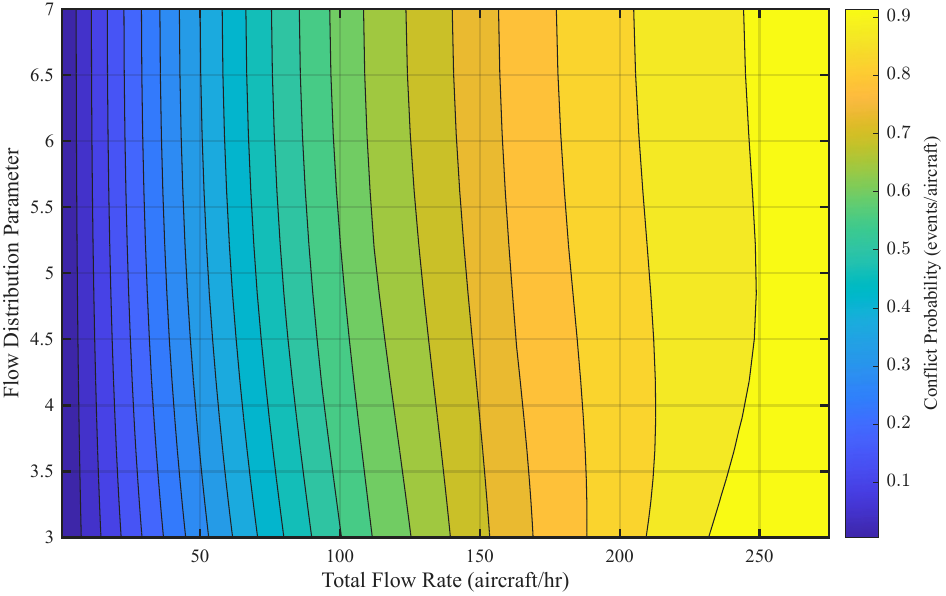}
    \caption{Cell Geometry 4 Conflict Probability as a function of flow rate and flow distribution.}
    \label{fig:PR_hex24}
\end{figure}

Given that a general airspace configuration will not have an equal flow rate from all origins, we investigate how conflict rates and probabilities change as the origin flow distribution changes. We maintain a degree of symmetry and linearly interpolate from equally distributing flows between the 3 by 3 routing geometry and the 7 by 7 routing geometry, while maintaining total flow rate. Near a Flow Distribution parameter of 3, aircraft will be constrained to entering and exiting on just the 3 by 3 routing geometry, while near a Flow Distribution parameter of 7, aircraft will be able to use the full 7 by 7 routing geometry equally. The conflict rates and probabilities of cell geometry 3 are shown in figures \ref{fig:Rate_hex13} and \ref{fig:PR_hex13}, respectively. As seen in the previous results, the scenarios with lower Flow Distribution Parameters correspond to lower collision probabilities at lower total flow rates, however, the trend becomes non-monotonic at higher flow rates when self interaction conflicts in the 3 by 3 routing geometry become dominant. The optimum Flow Distribution Parameter is just above 3, as this takes enough load off of the flows near maximum capacity while balancing the conflicts introduced by new flow interactions. The conflict rate and probabilities for cell geometry 4 are shown in figures \ref{fig:Rate_hex24} and \ref{fig:PR_hex24} respectively. The same trends appear as for geometry 3, however, the optimal Flow Distribution Parameter at high flow rates is much greater. This is attributed to the earlier onset of self interactions in geometry 4 due to the central routing and larger merge and crossing angles. 

These charts may also be used to set flow distributions that maximize routing flexibility given a required total flow rate and a workload constraint on the rate at which controllers can resolve conflicts in a given volume of airspace. If a workload driven limit on the conflict rate is imposed, airspace designers can follow a conflict rate isoline and maximize the Flow Distribution Parameter at given total flow rate. In an airspace design and workload allocation scenario, the relevant system level constraint may be conflict probability, such that controller workload is allocated on a per aircraft basis. In this case, airspace could be designed by maximizing the flow distribution parameter while observing constraints on the probability of conflict and the total flow rate.   

\section{Conclusion}
This paper introduced an analytical method for computing a bound on conflict rate and conflict probability in structured airspace flows based on renewal process models and geometry-dependent spacing constraints. Applications of the methodology allow for the investigation of tradeoffs between routing flexibility and total flow rates. Reduced flexibility routing options tend to result in lower conflict rates; however, they saturate earlier due to flow self-interaction. Routing geometry (direct vs central) significantly influences conflict rates through crossing angles and interaction concentration. 

The framework enables airspace designers to select flow distributions that balance throughput, routing flexibility, and controller workload constraints. Airspace may be optimized while observing workload constraints by designing along conflict rate iso-lines or allocating resources to equitably distribute conflict probability. Although this work was applied to level hexagonal cells with uniform origin flow rates, it is readily extensible to arbitrary sector geometries, altitude layering, and heterogeneous inter-arrival distributions. Fundamental limitations of the analysis required for computational tractability include the assumption of independence assumptions between origin flows, parametric inter-arrival models, and simplified deterministic kinematics. The methodology can be applied to optimize airspace composed of many coupled cells, dynamic flow management, and active flow control using timing and routing adjustments. By providing a tractable analytical link between flow structure and conflict dynamics, this work supports the systematic design of scalable and efficient airspace configurations.

\section*{Appendix}
We first compute the conflict probability bounds for the number of Case I arrivals in duration $s$, in which flows are from the same origin. The upper bounds on the arrival of 1, 2, and 3 aircraft in duration $s$ are derived in Eqs.\eqref{eq: 2_arrivalleq_mod_same}, \eqref{eq: 3_arrivalleq_mod_same}, and \eqref{eq: 4_arrivalleq_mod_same} respectively. The assumption that aircraft are already spaced with minimum inter-arrival time $s_0$ is used extensively.

\begin{equation}
\begin{aligned}
\mathbb{P}(S_0\leq s,S_0+S_1\geq s)&=\int_0^s f(t)\big[1- \int_t^s f(s-u) du \big]dt\\
&=\int_0^{s} \mathbf1_{t\geq s_0}f(t)[1- F(s-t) ]dt\\
&\leq (1-F(0))\int_{0}^{s} \mathbf1_{t\geq s_0} f(t)  dt\\
&\leq F(s) \\
\end{aligned}
\label{eq: 2_arrivalleq_mod_same}
\end{equation}

\begin{equation}
\begin{aligned}
\mathbb{P}(S_0+S_1 \leq s,S_0&+S_1+S_2>s)=\int_0^s f(t) \int_t^s f(u-t)\big[1-\int_u^s f(s-v)dv\big]  du dt \\ 
&=\int_0^{s} \mathbf1_{[t\geq s_0]}f(t) \int_t^s \mathbf1_{[u>t+s_0]}f(u-t)[1-F(s-u)] du dt \\ 
&=\int_0^{^{(s-s_0)^+}} f(t) \int_{t+s_0}^{s} f(u-t)[1-F(s-u)] du dt \\ 
&\leq\int_{s_0}^{(s-s_0)^+}  f(t) [1-F(s-s)] \int_{t+s_0}^{s} f(u-t) du dt \\ 
&\leq\int_{s_0}^{(s-s_0)^+}  f(t) F(s-t) dt \\ 
&\leq F(s-s_0)^+ \int_{s_0}^{(s-s_0)^+}  f(t)  dt \\ 
&\leq F^2(s-s_0)^+ \\ 
\end{aligned}
\label{eq: 3_arrivalleq_mod_same}
\end{equation}

\begin{equation}
\begin{aligned}
\mathbb{P}(&S_0+S_1+S_2 \leq s,S_0+S_1+S_2+S_3>s)=\\&=\int_0^s f(t) \int_t^s f(u-t)\int_u^s f(v-u)\big[1-\int_v^s f(s-w)dw\big] dv du dt \\ 
&=\int_0^{s}\mathbf1_{[t\geq s_0]} f(t) \int_t^s \mathbf1_{[u>t+s_0]}f(u-t)\int_u^s\mathbf1_{[v>u+s_0]} f(v-u)[1-F(s-v)] dv du dt \\ 
&=\int_{s_0}^{({s-2s_0})^+} f(t) \int_{t+s_0}^{(s-s_0)^+} f(u-t)\int_{u+s_0}^s f(v-u)[1-F(s-v)] dv du dt \\ 
&\leq\int_{s_0}^{({s-2s_0})^+} f(t) \int_{t+s_0}^{(s-s_0)^+} f(u-t)[1-F(s-s)]\int_{u+s_0}^s f(v-u) dv du dt \\ 
&\leq\int_{s_0}^{({s-2s_0})^+} f(t) \int_{t+s_0}^{(s-s_0)^+} f(u-t)F(s-u) du dt \\ 
&\leq\int_{s_0}^{({s-2s_0})^+} f(t) F(s-t-s_0) \int_{t+s_0}^{(s-s_0)^+} f(u-t) du dt \\ 
&\leq\int_{s_0}^{({s-2s_0})^+} f(t) F^2(s-t-s_0) dt \\ 
&\leq F^2(s-2s_0)^+ \int_{s_0}^{(s-2s_0)^+}  f(t)  dt \\ 
&\leq F^3(s-2s_0)^+ \\ 
\end{aligned}
\label{eq: 4_arrivalleq_mod_same}
\end{equation}

The general form inferred for Case I is as follows, where $J_n=\sum_{i=0}^nS_i$
\begin{equation}
\begin{aligned}
\mathbb{P}(&J_n \leq s,J_{n+1}>s)\leq F^{n+1}(s-ns_0)^+ .
\end{aligned}
\label{eq: arrivalleq_mod_same_infer}
\end{equation}

We similarly compute the probability bounds for the number of Case II arrivals during duration $s$. The upper bounds on the arrival of 1, 2, and 3 aircraft in duration $s$ are derived in Eqs.\eqref{eq: 2_arrivalleq_mod}, \eqref{eq: 3_arrivalleq_mod}, and \eqref{eq: 4_arrivalleq_mod} respectively, utilizing the same assumptions.

\begin{equation}
\begin{aligned}
\mathbb{P}(R\leq s,R+S_1\geq s)&=\int_0^s \phi(t)\big[1- \int_t^s f(s-u) du \big]dt\\
&=\int_0^{s} \phi(t)[1- F(s-t) ]dt\\
&\leq (1-F(0))\int_0^{s} \phi(t)  dt\\
&\leq (1-\Phi(s)) \\
\end{aligned}
\label{eq: 2_arrivalleq_mod}
\end{equation}

\begin{equation}
\begin{aligned}
\mathbb{P}(R+S_1 \leq s&,R+S_1+S_2>s)=\int_0^s \phi(t) \int_t^s f(u-t)\big[1-\int_u^s f(s-v)dv\big]  du dt \\ 
&=\int_0^{s} \phi(t) \int_t^s \mathbf1_{[u>t+s_0]}f(u-t)[1-F(s-u)] du dt \\ 
&=\int_0^{^{(s-s_0)^+}} \phi(t) \int_{t+s_0}^{s} f(u-t)[1-F(s-u)] du dt \\ 
&\leq\int_0^{(s-s_0)^+}  \phi(t) [1-F(s-s)] \int_{t+s_0}^{s} f(u-t) du dt \\ 
&\leq\int_0^{(s-s_0)^+}  \phi(t) F(s-t) dt \\ 
&\leq F(s) \int_0^{(s-s_0)^+}  \phi(t)  dt \\ 
&\leq F(s) (1-\Phi(s-s_0)^+) \\ 
\end{aligned}
\label{eq: 3_arrivalleq_mod}
\end{equation}

\begin{equation}
\begin{aligned}
\mathbb{P}(R+&S_1+S_2 \leq s,R+S_1+S_2+S_3>s)=\\&=\int_0^s \phi(t) \int_t^s f(u-t)\int_u^s f(v-u)\big[1-\int_v^s f(s-w)dw\big] dv du dt \\ 
&=\int_0^{s} \phi(t) \int_t^s \mathbf1_{[u>t+s_0]}f(u-t)\int_u^s\mathbf1_{[v>u+s_0]} f(v-u)[1-F(s-v)] dv du dt \\ 
&=\int_0^{({s-2s_0})^+} \phi(t) \int_{t+s_0}^{(s-s_0)^+} f(u-t)\int_{u+s_0}^s f(v-u)[1-F(s-v)] dv du dt \\ 
&\leq\int_0^{({s-2s_0})^+} \phi(t) \int_{t+s_0}^{(s-s_0)^+} f(u-t)[1-F(s-s)]\int_{u+s_0}^s f(v-u) dv du dt \\ 
&\leq\int_0^{({s-2s_0})^+} \phi(t) \int_{t+s_0}^{(s-s_0)^+} f(u-t)F(s-u) du dt \\ 
&\leq\int_0^{({s-2s_0})^+} \phi(t) F(s-t-s_0) \int_{t+s_0}^{(s-s_0)^+} f(u-t) du dt \\ 
&\leq\int_0^{({s-2s_0})^+} \phi(t) F^2(s-t-s_0) dt \\ 
&\leq F^2(s-s_0)^+ \int_0^{(s-2s_0)^+}  \phi(t)  dt \\ 
&\leq F^2(s-s_0)^+ (1-\Phi(s-2s_0)^+) \\ 
\end{aligned}
\label{eq: 4_arrivalleq_mod}
\end{equation}

The general form inferred for Case II is as follows, where $J_0 = R,\,\, J_{n\geq1}=R+\sum_{i=1}^nS_i$
\begin{equation}
\begin{aligned}
\mathbb{P}(&J_n \leq s,J_{n+1}>s)\leq F^{n}(s-(n-1)s_0)^+ (1-\Phi(s-ns_0)^+).
\end{aligned}
\label{eq: arrivalleq_mod_diff_infer}
\end{equation}

\begin{figure}
    \centering
    \includegraphics[width=0.9\linewidth]{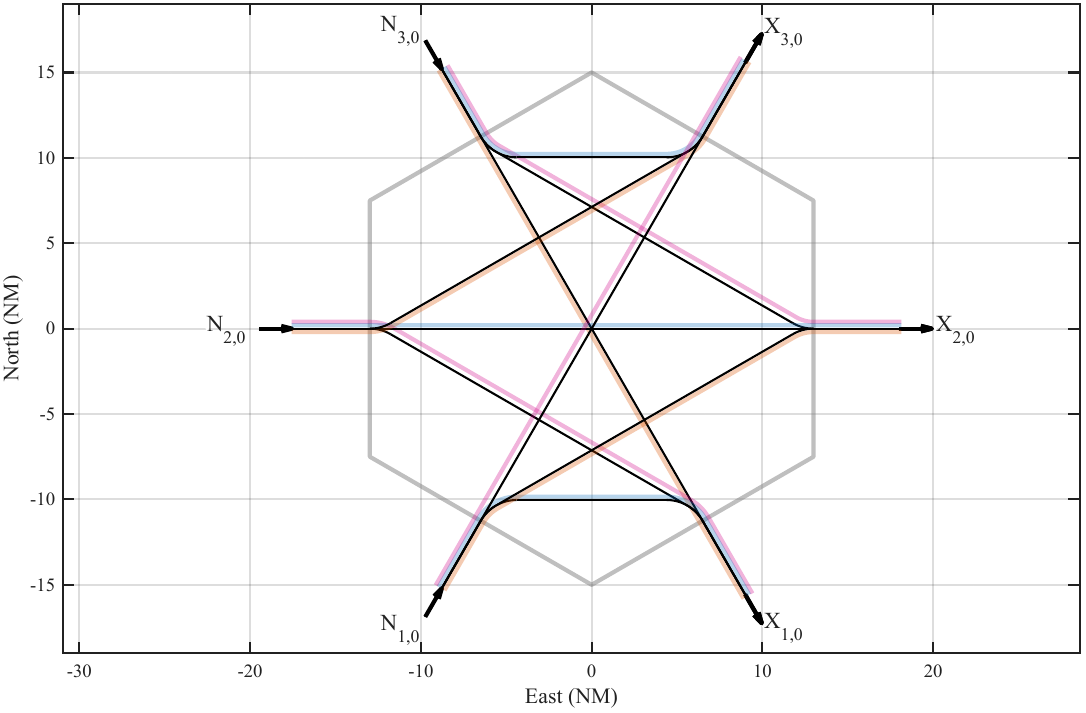}
    \caption{Cell 1 Routing Geometry}
    \label{fig:hex1_map}
\end{figure}

\begin{figure}
    \centering
    \includegraphics[width=0.9\linewidth]{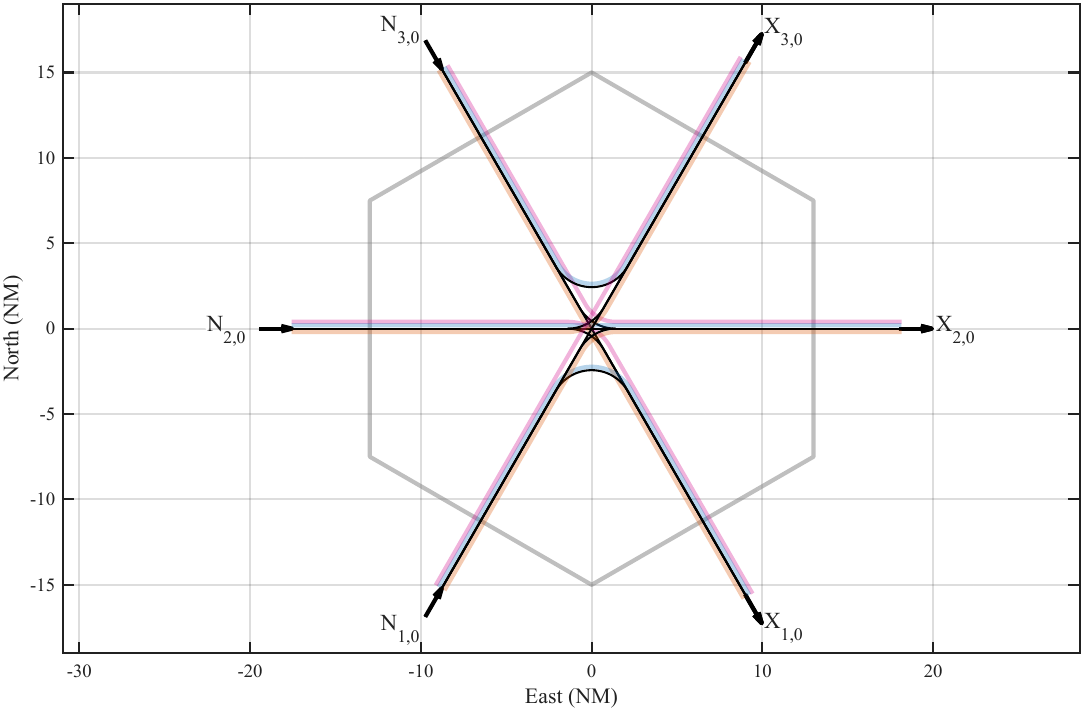}
    \caption{Cell 2 Routing Geometry}
    \label{fig:hex1_map}
\end{figure}

\begin{figure}
    \centering
    \includegraphics[width=0.9\linewidth]{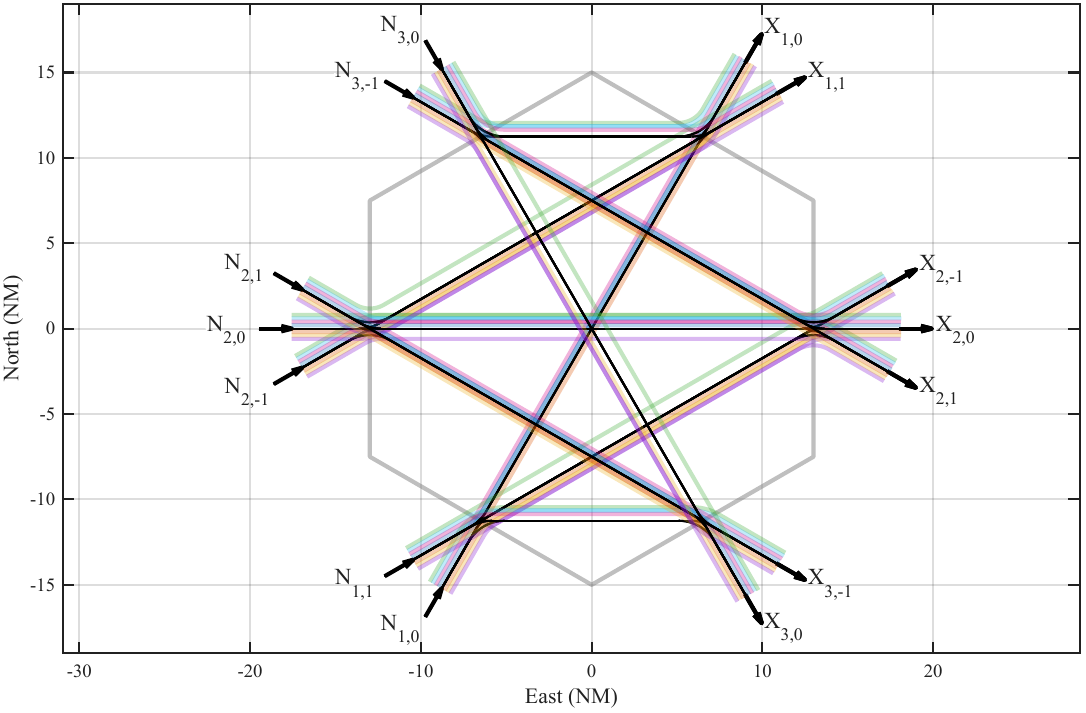}
    \caption{Cell 3 Routing Geometry}
    \label{fig:hex3_map}
\end{figure}

\begin{figure}
    \centering
    \includegraphics[width=0.9\linewidth]{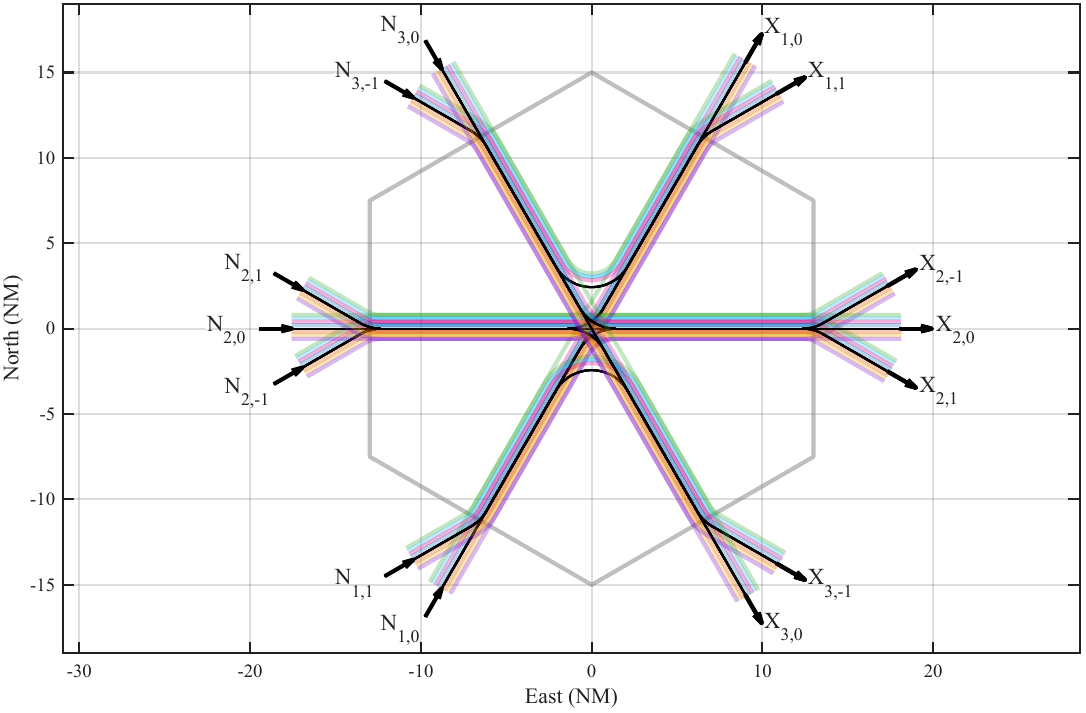}
    \caption{Cell 4 Routing Geometry}
    \label{fig:hex4_map}
\end{figure}

\begin{figure}
\centering
\begin{minipage}{.5\textwidth}
  \centering
  \includegraphics[width=1.0\linewidth]{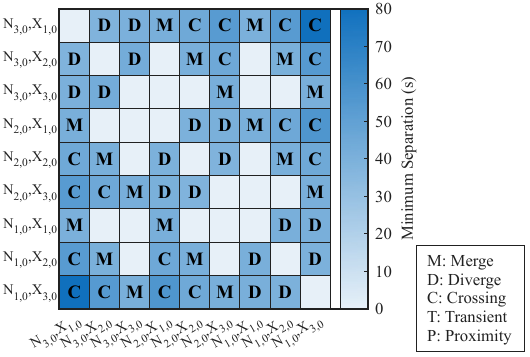}
  \captionof{figure}{Cell 1 Minimum Spacing}
  \label{fig:test1}
\end{minipage}%
\begin{minipage}{.5\textwidth}
  \centering
  \includegraphics[width=1.0\linewidth]{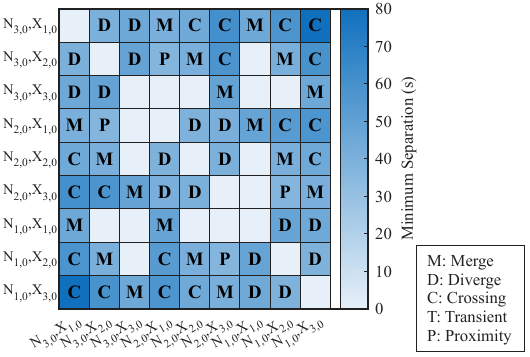}
  \captionof{figure}{Cell 2 Minimum Spacing}
  \label{fig:test2}
\end{minipage}
\end{figure}

\begin{figure}
   \centering
    \includegraphics[width=1.35\linewidth,angle=-90]{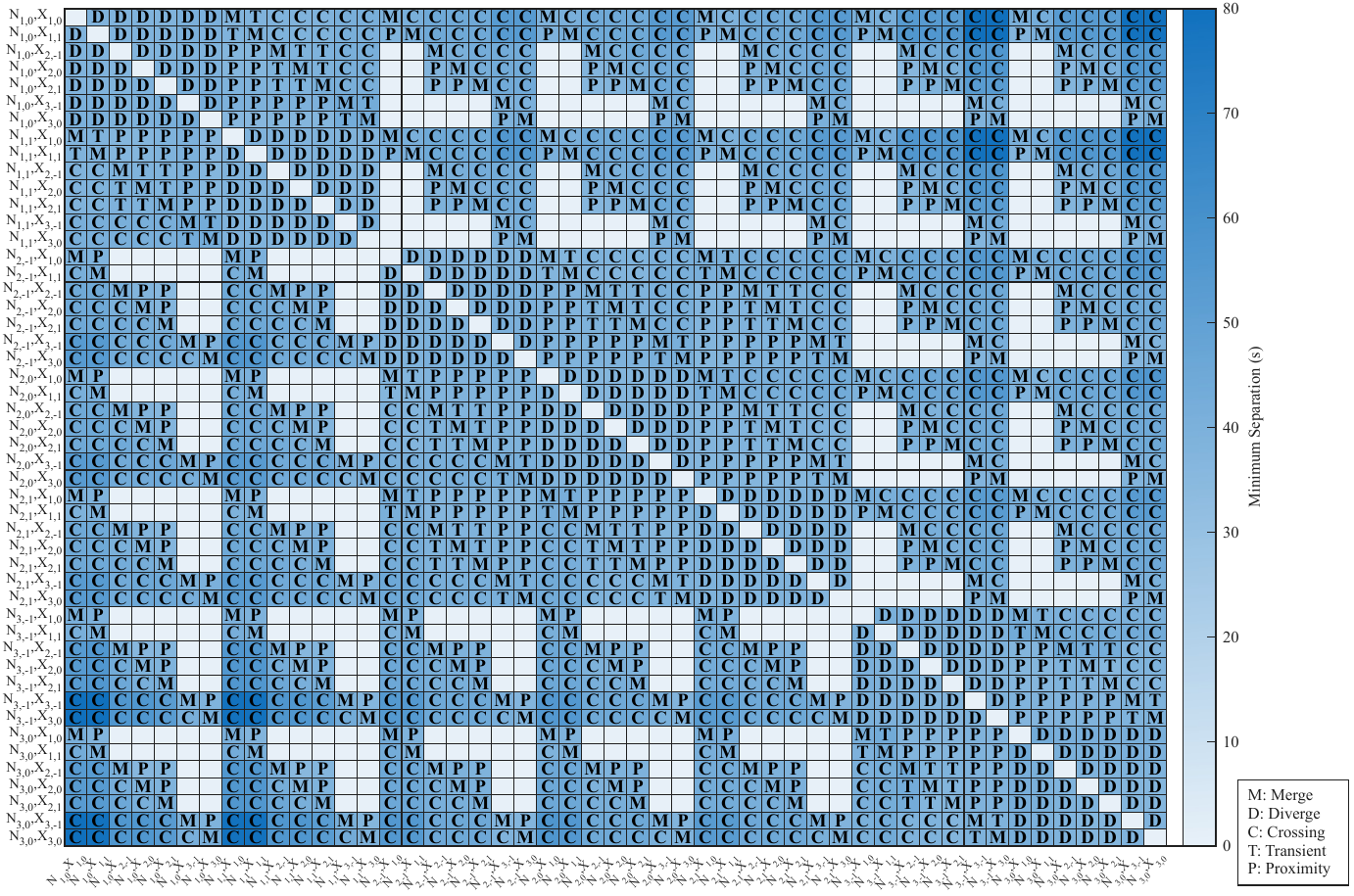}
    \caption{Cell 3 Minimum Spacing}
    \label{fig:hex3_map}
\end{figure}

\begin{figure}
   \centering
    \includegraphics[width=1.35\linewidth,angle=-90]{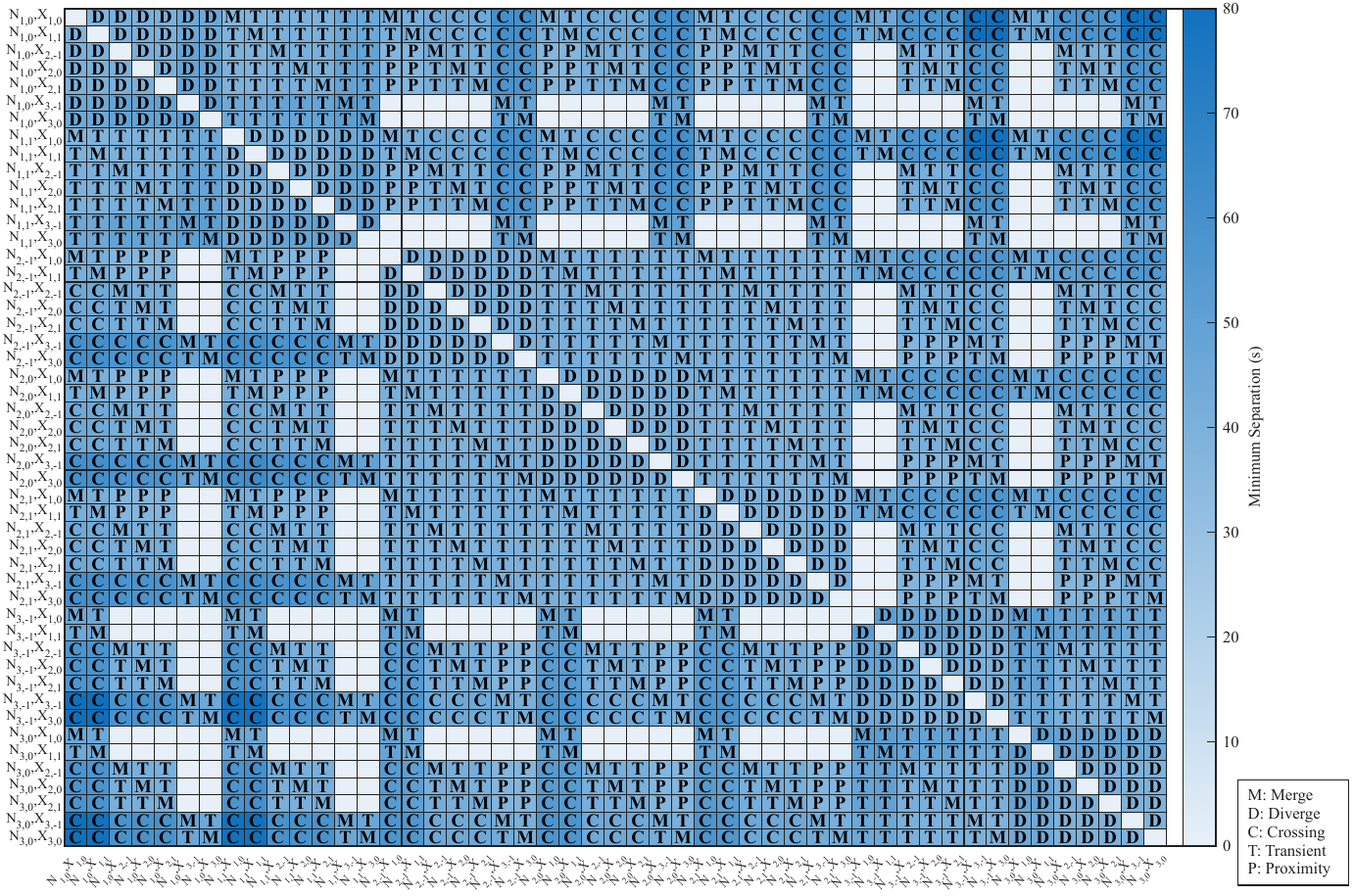}
    \caption{Cell 4 Minimum Spacing}
    \label{fig:hex4_map}
\end{figure}

\section*{Funding Sources}

This work was supported by the National Aeronautics and Space Administration (NASA) University Leadership Initiative (ULI) program under project “Autonomous Aerial Cargo Operations at Scale”, under grant No.80NSSC21M071 to the University of Texas at Austin. Any opinions, findings, conclusions, or recommendations expressed in this material are those of the authors and do not necessarily reflect the views of the project sponsor.

\section*{Acknowledgments}
The author would like to thank David Goldsman of the Georgia Institute of Technology for valuable discussions regarding the problem formulation and analytical development of this work.

\bibliography{ref}

\end{document}